\documentclass[12pt,preprint]{aastex}

\newcommand\beq{\begin{equation}} \newcommand\eeq{\end{equation}}

\def\mpc{{\rm Mpc}}
\def\a168{Abell~168}

\def\eg{e.g.,\ }
\def\x{\!\times\!}
\def\kms{$\rm{km\ s}^{-1}$}

\def\Msun{$M_{\odot}$}
\def\kpc{\,{\rm kpc}}

\received{-- ----} \accepted{-- ----}
\articleid
\date

\shorttitle{Chandra X-ray Analysis of Galaxy Cluster A168} 
\shortauthors{Yang et al.}

\begin{document}

{\title{Chandra X-ray Observations of Galaxy Cluster A168}

\author{Yanbin Yang\altaffilmark{1}, Zhiying Huo\altaffilmark{1}, Xu Zhou\altaffilmark{1}, Suijian Xue\altaffilmark{1}, Shude Mao\altaffilmark{2}, 
 Jun Ma\altaffilmark{1}, Jiansheng Chen\altaffilmark{1} 
}

\altaffiltext{1}{National Astronomical Observatories, Chinese Academy
of Sciences, Beijing, 100012, P. R. China }
\altaffiltext{2}{University of Manchester, Jodrell Bank Observatory,
Macclesfield, Cheshire SK11 9DL, UK}
\email{yyb@bac.pku.edu.cn
}

\begin{abstract}

We present {\it Chandra} X-ray observations of galaxy cluster 
A168 ($z=0.045$).
Two X-ray peaks with a projected distance of 676~kpc 
are found to be located close to two dominant galaxies, respectively.
Both peaks are significantly offset from the peak of the number density 
distribution of galaxies. 
This suggests that A168 consists of two subclusters,
a northern subcluster (A168N) and a southern subcluster (A168S).
Further X-ray imaging analysis reveals that
(1) the X-ray isophotes surrounding the two X-ray peaks 
are heavily distorted, 
(2) an elongated and continuous filament connects the two X-ray peaks. 
These suggest that strong interactions have
occurred between the two subclusters.
Spectral analysis shows that A168 has a mean temperature of $2.53 \pm 0.09$~keV
and a mean metallicity of $0.31\pm0.04Z_{\odot}$. The metallicity is
roughly a constant across the cluster but the temperature shows
some systematic variations.
Most X-ray, optical and radio properties of
A168 are consistent with it being an off-axis merger several Gyrs
after a core passage, although detailed numerical simulations are
required to see whether the observed properties, in particular the
significant offset between the optical and X-ray centers, can be
reproduced in such a scenario.
\end{abstract}


\section{INTRODUCTION\label{secINTRO}}
Merging of galaxy clusters is key to our understanding of 
the formation and evolution of not only the clusters themselves
\citep{2000MNRAS.318L..65F, 2002ApJ...565..867S, 2003MNRAS.344L..43F},
but also their member galaxies
in the dense and violent cluster environment \citep{2004cgpc.symp..278M}. 
Such merging processes can dissipate a 
vast amount of kinetic energy ($10^{63}\sim10^{64}$~erg)
via  different physical processes, such as
heating the gas, generating intense radio and X-ray radiations \citep{2001ApJ...563...95M}, 
and producing relativistic particles and magnetic fields
\citep{2001ApJ...551..160V, 2003ApJ...583..695G}. 
X-ray is a unique band for studying
clusters of galaxies, especially during mergers,
because X-ray studies directly probe the hot intracluster medium
and hence the dynamical properties of clusters of galaxies
\citep{2002ApJ...567L..27M,1997ApJS..109..307R,
1999ApJ...518..594R,1999ApJ...510L..15B,2001ApJ...561..621R}.
X-ray is hence a powerful tool for studying the formation and 
evolution of galaxy clusters.

In the present paper, we report the analysis of {\it Chandra} observations
of A168 which is located at
${01^{\rm h}15^{\rm m}06^{\rm s}}+00^\circ19^{\prime}12^{\prime\prime}$ J2000.0
\citep[Paper\,I hereafter]{2004ApJ...600..141Y} with type of I$\!$I$-$I$\!$I$\!$I \citep{1970ApJ...162L.149B}.
Two decades ago, A168 was observed by {\it Einstein} IPC, and two X-ray peaks
were reported \citep{1996AJ....111...42T}.
\citet{1992ApJ...397..430U} 
reported  
an unusual offset between the optical center and  the X-ray
center  as observed by {\it Einstein}. They
suggested that A168 was formed
by the collision of two approximately equal size subclusters.
Based on optical observations,
In Paper\,I 
we found that the elongated shape of the X-ray emission
is consistent with the spatial distribution of member galaxies. 
We also suggested that A168 possibly consist of two subclusters,
a northern subcluster (A168N) and a southern subcluster (A168S), based on
evidence from the velocity information and the luminosity
function of galaxies 
in the cluster. In this paper, we will use {\it Chandra} observations to study
A168. The {\it Chandra} data allows us to study this cluster in
un-precedented details and for the first time enables us to
separate clearly the emission from point sources and the diffuse
hot gas. Throughout this paper,  
we assume a cosmology with a matter density parameter
$\Omega_0=0.3$, a cosmological constant
$\Omega_{\Lambda,0}=0.7$, and a Hubble constant 
$H_0=75{\rm\,km\,s^{-1}Mpc^{-1}}$. In this
cosmology, the cluster has a luminosity distance $186\,\mpc$ and one
arcsecond corresponds to $\sim 0.81\kpc$.
All error ranges in this paper are quoted at the 90\% confidence level.

\section{OBSERVATIONS AND DATA REDUCTION \label{secDATA}}
A168 was observed twice by {\it Chandra} with the Advanced CCD Imaging
Spectrometer (ACIS-I).
The first observation, called A168\_OFFSET1 (OF1), started on 
2002 January 5. OF1 was centered on
${01^{\rm h}14^{\rm m}52^{\rm s}\!.30}+00^\circ23^{\prime}58^{\prime\prime}$
(J2000.0)
with an exposure time of 41.12\,ks.  
The second observation
(A168\_OFFSET2, hereafter OF2), 
centered on ${01^{\rm h}15^{\rm m}25^{\rm s}\!.00}
+00^\circ16^{\prime}15^{\prime\prime}$, was performed on November 1
in the same year with an exposure time of 38.11\,ks.
Both observations were telemetered in the VFAINT mode.
Following the standard steps we processed
the raw data with {\it Chandra} Interactive Analysis of Observations
({\tt CIAO}, v3.0) and Calibration Database ({\tt CALDB}, v2.23)%
\footnote{{\tt CIAO} and {\tt CALDB} can be found at website of http://cxc.harvard.edu.}.

The data reduction for the {\it Chandra} observations of A168 offers some challenges.
First, A168 is a faint and extended source with an angular size
of at least 12$^{\prime} \x 18^{\prime}$,
occupying almost the whole CCD.
It is hard to find a clean area to determine the background reliably.
Accordingly, we have to take the blank-sky 
\citep{2001ApJ...562L.153M,2003ApJ...583...70M}, 
as the background in the spectral analysis.
We also corrected for the degradation of the low energy quantum efficiency
which is particularly pronounced at energies below 1~keV.
Second, there is an overlap region between OF1 and OF2.
It is easy to analyze the spectra in this region by extracting the spectra
separately and fitting them simultaneously. However, we found from
experimenting that it is not straightforward to add the images of
the two observations together without creating some artifacts.
We finally decided to take a simpler approach by considering only 
one observation in the overlap region.
We then correct the exposure map for the two pointings and combine
them to obtain the final image (see Figure~\ref{figIMGreg}).
Third, because A168 is faint and extended in the X-ray, the photon
noise is high at the full resolution of {\it Chandra}.
Therefore we binned the image in $8 \x 8$  ($3.94^{\prime\prime}$ 
by $3.94^{\prime\prime}$) pixels.
The binned image is then cropped to a smaller area with 
an effective size of $512 \x 512$ pixels.

\section{RESULTS\label{secRESULTS}}

\subsection{Imaging \label{secIMG}}
Figure~\ref{figIMGxray} shows the exposure map corrected 
and adaptively smoothed\footnote{We refer the {\tt csmooth} 
algorithm to http://cxc.harvard.edu/ciao/ahelp/csmooth.html. }
image of A168 with point sources excluded.
The 7-pixel-width gaps between CCD chips and the CCD boundaries are cut
for their high unreliability.
The image is smoothed by requiring the signal-to-noise ratio (SNR)
to equal 6 with a Gaussian kernel; the SNR is defined as
${\rm SNR}(r)=({C - A C_{\rm bg}})/{\sqrt{(\sqrt{C})^2+A\sigma_{\rm
      bg}^2}}$, where $C$ is the total count within an
aperture $r$ (in units of pixel),
which is taken to be the scale of the smoothing kernel, 
$A$ is area, i.e., the number of pixels within the aperture, 
$C_{\rm bg}$ is the mean counts of background, 
and  $\sigma_{\rm bg}$ is the variation of background.
$C_{\rm bg}$ and $\sigma_{\rm bg}$ are estimated from the three box regions 
in Figure~\ref{figIMGxray}.
The typical smoothing scale of $\sim6$ pixels ($\sim$ 24\arcsec).
In the most luminous region the scale is $2\sim3$ pixels.
The maximum smoothing scale is taken to be 10 pixel,
resulting in low SNRs ($\sim 1$-$2$) in regions where the photon counts
are close to the
background count rate.

Clearly, Figure~\ref{figIMGxray} shows that 
A168 in X-ray has two X-ray peaks, peak A and peak B.
These two X-ray peaks have an angular separation of
$\sim 13'\!.6$, corresponding to a projected distance of 
676 kpc. We also
notice that the X-ray emissions around the two X-ray peaks are
heavily distorted. Figure~\ref{figIMGxray} 
also shows an elongated filamentary structure
connecting these two peaks with a prominent clump in the middle.
To show the significance of the filament, we study the photon counts
in the dashed rectangle shown in Figure~\ref{figIMGxray}. 
We divide the $x$-axis into 20 bins, and count the photon events along
the $y$-axis for each bin with the background counts subtracted. 
To allow for the gap between CCDs, we normalize the counts by the total
number of pixels in each bin. The
result is shown in Figure~\ref{figIMGxray}$a$,
which suggests that emission from the filament is significant.
We repeat the same procedure in the $y$-direction (show in
Figure~\ref{figIMGxray}$b$), the photon counts indicate that the filament 
extends continuously from the south to north. 

For further analysis, we divide the emission into three energy bands.
Figure~\ref{figIMGcolor} shows the contours of  the
soft (0.3-1.5 keV), medium (1.5-2.5 keV) and hard (2.5-10 keV) bands
together with a 3-color image combined linearly from these bands.
The color image shows some structures which are suggestive of
slight temperature variations across A168; we address this question
in the next subsection. It is worth noting that there is a small blue clump 
located to the northeast of peak B
(see Figure~\ref{figIMGcolor}$c$ and the contours of the hard band).

\subsection{Spectrum and Temperature Map\label{secSPEC}}

The X-ray emission from the hot gas in clusters of galaxies is 
thought to be from thermal bremsstrahlung, which is usually
described by a 
{\tt MEKAL} model \citep{1993A&AS...97..443K,1995ApJ...438L.115L}. 
We therefore perform our spectral analysis using a single temperature 
{\tt WABS(MEKAL)} model in the {\tt XSPEC} package (v11.2.0).  The
absorption column density is fixed to the Galactic value,
${N_{\rm H}}=3.4\times10^{20}{\rm cm}^{-2}$ \citep{1990ARA&A..28..215D},
as the intrinsic absorption in the cluster is likely to be small.
All the following spectra are analyzed in the 0.5-0.8~keV energy range
with the point sources removed.
The corresponding background spectra of source spectra are extracted 
separately from the corresponding blank-sky regions on the CCD.

We first investigate the global temperature and metallicity of A168.
We define an overall region by the boundary of the
thin-solid boxes in Figure~\ref{figIMGreg}. From the spectral
fitting, the mean temperature and metallicity are found to be  
$2.53\pm0.09$~keV and $0.31\pm0.04Z_{\odot}$, 
respectively. The mean temperature is 
in agreement with previous work by \citet{1993ApJ...412..479D}
which gives $kT=2.6^{+ 1.1}_{- 0.6}$~keV. 
The total luminosity is $(0.39\pm0.01)\x10^{44}$ ergs/s over 
the 0.01-80 keV energy range. The temperature and luminosity
are consistent with the modest mass of cluster (see Table~\ref{tblA168NS}).
\citet{1997MNRAS.292..419W} gave a total luminosity of 
$(1.13\pm0.03)\x10^{44}$ ergs/s in the same energy band 
but for a cosmology with
$H_0=50{\rm\,km\,s^{-1}Mpc^{-1}}$, $q_0=0.5$ and a metallicity of
$0.4 Z_\odot$ (by assumption). If we convert our results to
their cosmology, we find that
$L_{\rm x}\sim0.78 \times 10^{44}$ ergs/s for the diffuse emission and 
$L_{\rm x}\sim0.23\times 10^{44}$ ergs/s for the point sources; the 
total luminosity is $1.01 \times  10^{44}$ ergs/s, in good
agreement with their value. Notice that 
the point sources contribute about 23\% of the total luminosity.
Without the resolutions of {\it Chandra}, these point sources will be
difficult to separate from the diffuse X-ray emission.


We next define a main body of A168 and divide it into 7 regions that are
marked as region A-G from south to north 
in Figure~\ref{figIMGreg} (thick-dashed boxes).
The spectra of the regions are extracted separately.
The photon count in each region is about 4000-5000 cts on average.
Region C and D has about twice the counts because the spectra are extracted
from both observations (OF1 and OF2).
The fitting results are listed in Table~\ref{tblT7} and plotted in
Figure~\ref{figTA7}. As can be seen, 
the metallicity is consistent with a constant from south to north.
while the fitted temperature appears to vary from south to north with
region C having the highest temperature.
To further investigate the temperature variation,
we increase the region under study
and divide it into 35 smaller boxes
as shown in Figure~\ref{figIMGreg} (thin-solid boxes).
The average count in each box is about 1300
(in the overlap region, the counts may be twice as large),
which are too few to constrain both the temperature and metallicity reliably.
As the metallicity is consistent with a constant according to 
the above analysis, we fix it to the mean value, $0.31\,Z_{\odot}$
(see the above text or Table \ref{tblT7}), and concentrate on
 the temperature variation.
The fitting results are listed in Table \ref{tblT35}. 
A temperature map (shown in Figure~\ref{figTmap})
is obtained through interpolation 
using the Kriging method%
\footnote{A function in the Winsurf 7.0 software \citep{1990IJGICS...4..313O}.
}.
We performed various $\chi^2$ tests 
to check the significance of the temperature variations. It appears
that region G has a significant lower temperature than other regions
(A-E, see Fig. \ref{figTA7})
while other temperature variations have less statistical significance.

\section{DISCUSSION\label{secDISCUSSION}}

In Sect.\,1 we mentioned that A168 is thought to be 
formed by the collision of two subclusters. This was first suggested from
the unusually large offset  between its X-ray and optical centers 
\citep{1992ApJ...397..430U}.
In paper\,I we found evidence for the existence of two subclusters
(A168N and A168S) in the optical from
the luminosity function and velocity distribution of galaxies in the cluster. 
As we argue below, {\it Chandra} X-ray observations further strengthen
the conclusion.

In Figure~\ref{figoptxray}, we show a comparison of the optical and
X-ray emissions.
Peak A is associated with the cD galaxy UGC~797, the brightest member.
Peak B is very close to an elliptical galaxy GIN~061,
the second brightest member galaxy. Table \ref{tblA168NS} 
collects the relevant properties of these two peaks
and their associated galaxies and subclusters. 
In the SLOAN $r$-band, the two associated galaxies have a luminosity
ratio of $2.5:1$. However, in the radio, UGC~797 is roughly a
factor of ten fainter than GIN~061  at 1.4 GHz \citep{1998AJ....115.1693C}.
The absolute magnitude of GIN~061 and its radio properties 
suggest that it may be a dominant galaxy of its local region. 

D/cD galaxies are commonly found in the evolved clusters or groups
\citep{1975ApJ...199..545M,1977ApJ...211..309A},
and they usually coincide with the X-ray peak in the
center of regular relaxed clusters of galaxies  
\citep{1965ApJ...142.1364M,1970ApJ...162L.149B,1979ApJ...234L..21J,1983ApJ...274..491B,2001AJ....122.2858O}.
However, in A168, while both dominant galaxies are clearly associated with
X-ray peaks, both show
significantly offset from the surface number
density peak of the galaxy distribution. This
is illustrated in Figure~\ref{figPaperI}, where 
we superpose the X-ray emission on the contours of the 
distribution of member galaxies (cf.\ Paper\,I). 
The number density peak of galaxies is located almost in the middle of
the two X-ray peaks. The offset between the optical and
X-ray centers clearly indicates that A168 is not a relaxed cluster.
Notice that the projected distance between the two dominant galaxies is
676\,kpc, comparable to the scale of clusters. The large
separation implies that these two galaxies must have formed  
separately in different clusters and hence have distinct evolution histories.

The luminosities of the two X-ray peaks are comparable (see Table 3),
with a luminosity ratio of $L_{\rm x}^{\rm A}/L_{\rm x}^{\rm B} \approx 1.5$.
Between the two X-ray peaks,
{\it Chandra} observations reveals an elongated and continuous filament
connecting the two, suggesting that
strong interactions have occurred between the two clumps.
The existence of the two peaks (A and B) in both
the X-ray and radio, and the X-ray filament connecting the two peaks
strongly suggest that A168 consists of
two interacting subclusters, A168N and A168S.

Table \ref{tblA168NS} collects some dynamical properties of A168N and A168S.
The mean velocity difference is only $264\pm142$ \kms, which is small
compared with the observed radial velocity dispersions ($\sigma_r \sim
600$ \kms), suggesting that the main bulk motion between these two
subclusters is perpendicular to the line of sight.
 In Paper\,I  we showed that the two components are  
gravitationally bound at the 92\% confidence level under a linear two-body
model \citep{1982ApJ...257...23B}.  
This result further supports the interaction between the 
two subclusters.  

From contours of image (Figure~\ref{figoptxray}), one can see that
the innermost regions of the two X-ray peaks are roughly symmetric,
but the outer contours surrounding two X-ray peaks are distorted,
but the distortions
 are not exactly along the line connecting the two X-ray peaks.
As the cluster temperature is not very high ($kT \sim 2.53$~keV), the
potentials of the two subclusters are not very deep, so their
outer contours may be easily distorted in the tidal interaction.
Furthermore, the filament and the prominent clump along it (see Fig. 5)
can be interpreted as the tidal-stripped gas from two subclusters.
These structures may also reflect the distribution of unrelaxed dark matter, 
similar to the ridge discovered in A1367
\citep{2002ApJ...576..708S}.}
\citet{1998ApJ...496..670R} and \citet{2001ApJ...561..621R} have studied
off-axis mergings of two comparable clusters.
Comparing with their results (cf.\ their Figs.\ 4-9), the X-ray
emission morphology of A168 resembles
that of an off-axis merger with a mass ratio from $1:1$ to
$1:3$ several Gyrs after a core passage.

We do not detect other significant changes of gas dynamics, 
such as shocks, strong temperature variations, as expected during 
merging \citep{1997ApJS..109..307R,2001ApJ...561..621R}.
Moreover, we do not find any radio lobes or haloes 
\citep[see][Figure~2$c$]{1999ApJS..120..147M}
which are considered to be related
to the merging or cooling activities of clusters of galaxies 
\citep{2000MNRAS.318L..65F, 2002ApJ...575..764F}. 
Furthermore, if the merger is ongoing, enhanced and 
concentrated blue galaxies are expected to be observed 
\citep{1988IAUS..130..311D}, but 
\citet{1996AJ....111...42T} concluded that there are 
no significant blue galaxies related to the merger.
The lack of activities seem to be consistent with the view
that the last major collision (core passage) has occurred 
several Gyrs ago and the merging signatures 
have `decayed'. However, it is not clear whether this timescale is
correct. We can estimate the timescale 
to be $\sim 0.6 $ Gyr
if we take the distance of two X-ray peaks to be 676~kpc, and assume
a colliding velocity of $\sim \sigma_r \sim 600$~km~s$^{-1}$.
Such a short time scale
would be difficult to reconcile with the lack of strong merging 
signatures as some
simulations show that the remnant signatures may last several Gyrs
\citep{1997ApJS..109..307R,2001ApJ...561..621R}. 
The most puzzling feature of A168 remains the fact that
the density peak of the galaxy number distribution seems
to be in the middle of the X-ray peak, this appears to be rare
in numerical simulations where the X-ray peaks
and the dark matter density peaks (and presumably the galaxy
distribution) usually coincide \citep{1998ApJ...496..670R,2001ApJ...561..621R}.

\section{SUMMARY}

We have analyzed the data from {\it Chandra} observations of A168. 
The mean temperature and metallicity 
are $2.53 \pm 0.09$~keV and $0.31\pm0.04Z_{\odot}$, respectively. 
The total luminosity of A168 is $(0.39\pm0.01)\x10^{44}$ ergs/s over the
0.01-80 keV energy range.
We divided the
system into many regions and studied their images and
spectral properties. Our main findings are as follows:
\begin{enumerate}
\item Two X-ray peaks with a projected distance of 
676~kpc are resolved unambiguously. 
The luminosity within an aperture of $49\arcsec$ and over
0.5-8.0 keV energy range is 
$(0.62\pm0.05)\x10^{42}$ ergs/s for peak A and 
$(0.41\pm0.04)\x10^{42}$ ergs/s for peak B. 
\item A significant and continuous filament 
with a prominent clump in the middle
is detected between the two X-ray peaks.
\item No significant metallicity gradient is seen across the cluster.
There is some variation in the temperature with region C 
having the highest temperature and region G
having a significantly lower temperature than other regions
(cf. Fig. \ref{figTA7} and \ref{figTmap}).
\end{enumerate}
The radio, optical and X-ray data for
A168 strongly suggest the existence of two subclusters, A168N and A168S. 
The X-ray morphology is consistent with it being an off-axis
merger of two comparable subclusters several Gyrs after a core
passage. But it is not clear whether
the lack of other merging signatures in the
X-ray (such as strong shocks) is consistent with this scenario.
A168 may be a very unusual system as 
the significant offset between the X-ray and optical centers
is rare.  It will be a challenge to see whether numerical simulations
can reproduce the observed X-ray properties, 
down to a spatial resolution of
$\sim0.4$~kpc at the full resolution of {\it Chandra}.
More future observations of {\it Chandra} or {\it XMM-Newton}
and detailed numerical simulations will
shed further insights on the nature of the system.

\acknowledgments 
We would like to thank the referee for valuable suggestions.
We thank Drs. Xiangping Wu, A.C. Fabian and Bo Qin 
for their valuable suggestions. 
We are also grateful to
Drs. Haiguang Xu, Jiasheng Huang, Xiaofeng Wang,
Zhengyu Wu, and especially Mr.\ Albrecht R\"udiger,
for helpful discussions.
This research has made use of 
the {\it Chandra} X-ray databases and the NASA/IPAC
Extragalactic Database (NED).
This work is
supported by the National Key Base Sciences Research Foundation under
the contract TG1999075402 and is also supported by the Chinese
National Science Foundation (NSF) under the contract No.~10273007.


\begin{table}[ht]
{\footnotesize
\caption{\label{tblT7}Temperatures and metallicities for the 7 regions
  (defined in Fig. 1) and the whole region.}
\vspace {0.5cm}
\begin{tabular}{cccc}
\tableline \tableline
Region
&Temperature (keV)
&Metallicity ($Z_{\odot}$) & $\chi^2$/d.o.f. \\
\tableline 
 A
&$ 2.48^{+ 0.29}_{- 0.27}$
&$ 0.24^{+ 0.14}_{- 0.11}$ & 168/122\\
 B			            
&$ 2.80^{+ 0.32}_{- 0.30}$          
&$ 0.28^{+ 0.17}_{- 0.13}$ & 182/158\\
 C			            
&$ 3.09^{+ 0.23}_{- 0.21}$          
&$ 0.39^{+ 0.11}_{- 0.09}$ & 381/333\\
 D			            
&$ 2.90^{+ 0.19}_{- 0.16}$          
&$ 0.24^{+ 0.08}_{- 0.07}$ & 406/341\\
 E			            
&$ 2.74^{+ 0.14}_{- 0.15}$          
&$ 0.35^{+ 0.09}_{- 0.07}$ & 349/281\\
 F			            
&$ 2.59^{+ 0.21}_{- 0.21}$          
&$ 0.43^{+ 0.15}_{- 0.12}$ & 220/174\\
 G			            
&$ 1.75^{+ 0.10}_{- 0.10}$          
&$ 0.42^{+ 0.13}_{- 0.10}$ & 210/142\\
 whole region
&$ 2.53^{+ 0.09}_{- 0.09}$
&$ 0.31^{+ 0.04}_{- 0.04}$ & 1569/1006\\           
\tableline
\end{tabular}\\
}
\end{table}

\newpage

\begin{table}[ht]
{\footnotesize
\caption{\label{tblT35}Temperatures (in units of {\rm keV}) for the 35 boxes
  as indicated in Figure~\ref{figIMGreg}; the metallicity has been fixed
to 0.31$Z_\odot$ in the spectral fitting.}
\vspace {0.5cm}
\begin{tabular}{c|ccccc}
\tableline \tableline
  & C1 & C2 & C3 & C4 & C5 \\
\tableline 
R1&$ 3.15^{+ 4.53}_{- 1.49}$&$ 3.56^{+ 2.75}_{- 1.21}$&$ 1.60^{+ 0.10}_{- 0.10}$&$ 1.61^{+ 0.18}_{- 0.18}$&$ 2.37^{+ 1.97}_{- 0.82}$ \\
R2&$ 2.32^{+ 0.99}_{- 0.43}$&$ 2.15^{+ 0.46}_{- 0.34}$&$ 2.74^{+ 0.33}_{- 0.27}$&$ 2.20^{+ 0.32}_{- 0.22}$&$ 1.75^{+ 0.24}_{- 0.18}$ \\
R3&$ 2.42^{+ 0.35}_{- 0.22}$&$ 2.50^{+ 0.24}_{- 0.21}$&$ 2.75^{+ 0.20}_{- 0.20}$&$ 2.67^{+ 0.25}_{- 0.26}$&$ 2.72^{+ 0.54}_{- 0.44}$ \\
R4&$ 2.77^{+ 0.42}_{- 0.39}$&$ 3.12^{+ 0.44}_{- 0.38}$&$ 2.73^{+ 0.24}_{- 0.24}$&$ 2.93^{+ 0.29}_{- 0.25}$&$ 2.98^{+ 0.48}_{- 0.35}$ \\
R5&$ 2.66^{+ 0.47}_{- 0.39}$&$ 3.52^{+ 0.54}_{- 0.47}$&$ 2.77^{+ 0.26}_{- 0.25}$&$ 2.85^{+ 0.46}_{- 0.29}$&$ 3.37^{+ 0.71}_{- 0.53}$ \\
R6&$ 2.92^{+ 1.16}_{- 0.74}$&$ 2.97^{+ 0.71}_{- 0.45}$&$ 2.21^{+ 0.35}_{- 0.24}$&$ 2.98^{+ 0.68}_{- 0.43}$&$ 1.82^{+ 0.31}_{- 0.20}$ \\
R7&$ 4.08^{+ 3.81}_{- 1.39}$&$ 4.13^{+ 2.69}_{- 1.29}$&$ 2.46^{+ 0.37}_{- 0.30}$&$ 2.45^{+ 0.48}_{- 0.31}$&$ 2.36^{+ 1.67}_{- 0.67}$ \\  
\tableline
\end{tabular}\\
}
\end{table}

\newpage

\begin{table}[ht]
{\footnotesize
\caption{\label{tblA168NS}Statistics of A168N and A168S.}
\vspace {0.5cm}
\begin{tabular}{crccc}
\tableline \tableline
 & & Peak A & Peak B & References \\ \tableline
\multicolumn{2}{l}{(Ra., Dec.) J2000.0}
& ${01^{\rm h}14^{\rm m}57^{\rm s}\!.81}+00^\circ25^{\prime}45^{\prime\prime}$ 
& ${01^{\rm h}15^{\rm m}15^{\rm s}\!.92}+00^\circ12^{\prime}54^{\prime\prime}$
& \\
 \multicolumn{2}{l}{Projected distance from A to B} 
& \multicolumn{2}{c}{ $676$$h^{-1}_{0.75}$~kpc} &  \\
\multicolumn{2}{l}{$L_{\rm x}$ ($10^{42}$ ergs/s)\tablenotemark{a}} 
& $0.62\pm0.05$  & $ 0.41\pm0.04$ & \\
\multicolumn{2}{l}{Asocciated galaxy} & UGC~797 & GIN~061 & (1) \\
 & $m_r$  & 13.58 & 14.59 & (2)\\
 & $M_r$  & $-22.77$ & $-21.76$ & \\
 & Radio flux at 1.4 GHz (mJy) 
    &   $4.5\pm0.6$  
    & $43.0\pm1.4$& (3) \\
 & Diameter (arcmin)  &  0.9 $\x$ 0.9  & 0.58 $\x$ 0.5 & (1) \\
 & $v_{\rm r}$ (\kms) & $13460\pm26$ & $13553\pm30$ & (1)\\
\multicolumn{2}{l}{Associated subcluster} & A168N & A168S & (4) \\
&$v_{\rm r}$ (\kms)
&$ 13065 \pm   113$ 
&$ 13329 \pm   86 $ & (4) \\
& $\Delta v_{\rm r}$ (\kms) 
& \multicolumn{2}{c}{ $264\pm142$} & (4) \\
& $\sigma_{\rm r}$ (\kms)
&$ 564  \pm     90 $
&$ 613  \pm     56 $
& (4) \\ 
& $M$ ($10^{14}$\,\Msun)
& $2.1  \pm    0.7$ 
& $2.5  \pm    0.5$
& (4) \\
\tableline
\end{tabular}\\
\tablenotetext{\rm a}{The flux is calculated over the 0.3-8.0 keV energy
  range within an aperture of 49\arcsec ($\sim40$ kpc), as indicated by two circles in Figure~\ref{figIMGxray}.}

References: 
(1) NASA/IPAC Extragalactic Database (NED); 
(2) Sloan Digital Sky Survey (SDSS), Data Release 2 (DR2); 
(3) \citet{1998AJ....115.1693C};
(4) \citet{2004ApJ...600..141Y}.
}
\end{table}

\newpage

\begin{figure}
\epsscale{0.8}
\plotone{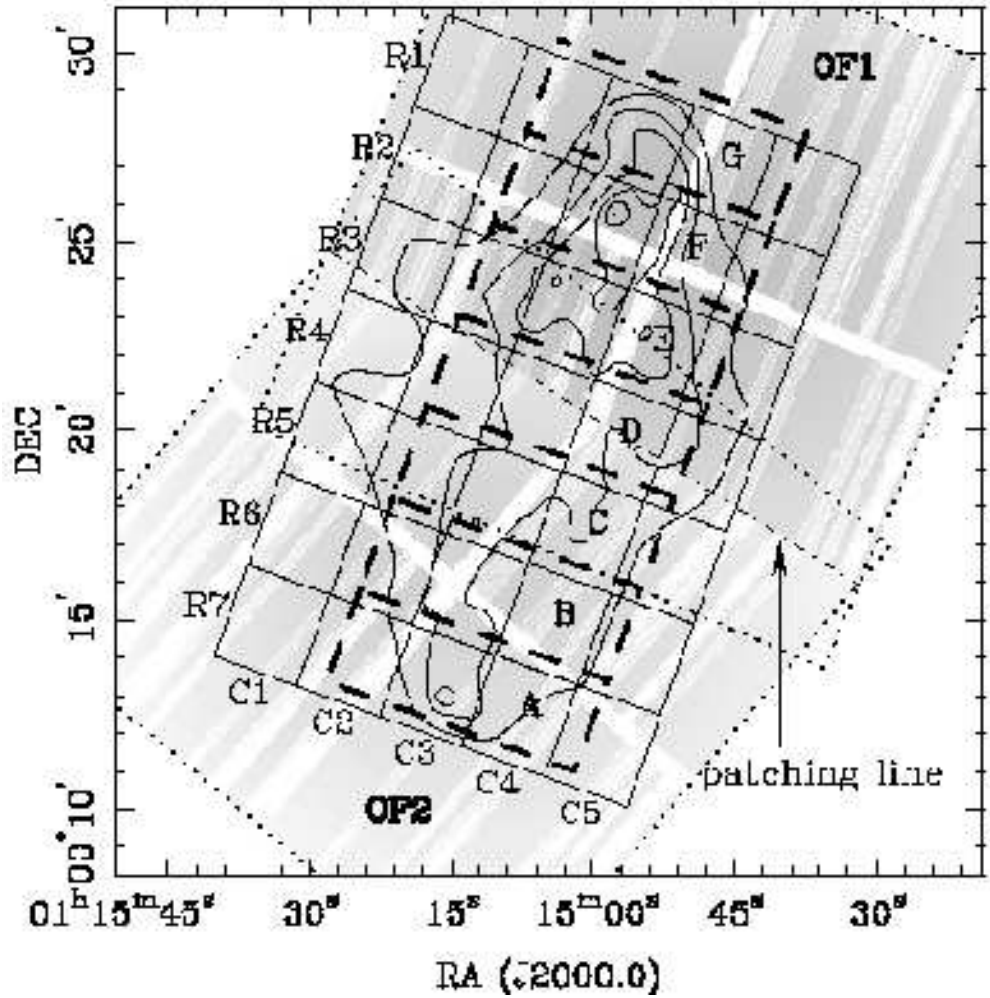}
\caption{
Dotted lines show the CCD field of view for each observation, OF1 and OF2.
The background image is a mosaic exposure map with CCD gaps excluded.
This image also demonstrates the mosaic style of the
ACIS-I image of A168 along the patching line.
Thin-solid lines and thick-dashed lines indicate 
the regions used for spectral analysis (see Sect.~\ref{secSPEC} and
Tables 1 and 2).
A rough contour of smoothed image (Figure~\ref{figIMGxray})
is shown for the comparison with regions of spectral analysis.
\label{figIMGreg}}
\end{figure}

\newpage

\begin{figure}
\epsscale{0.8}
\plotone{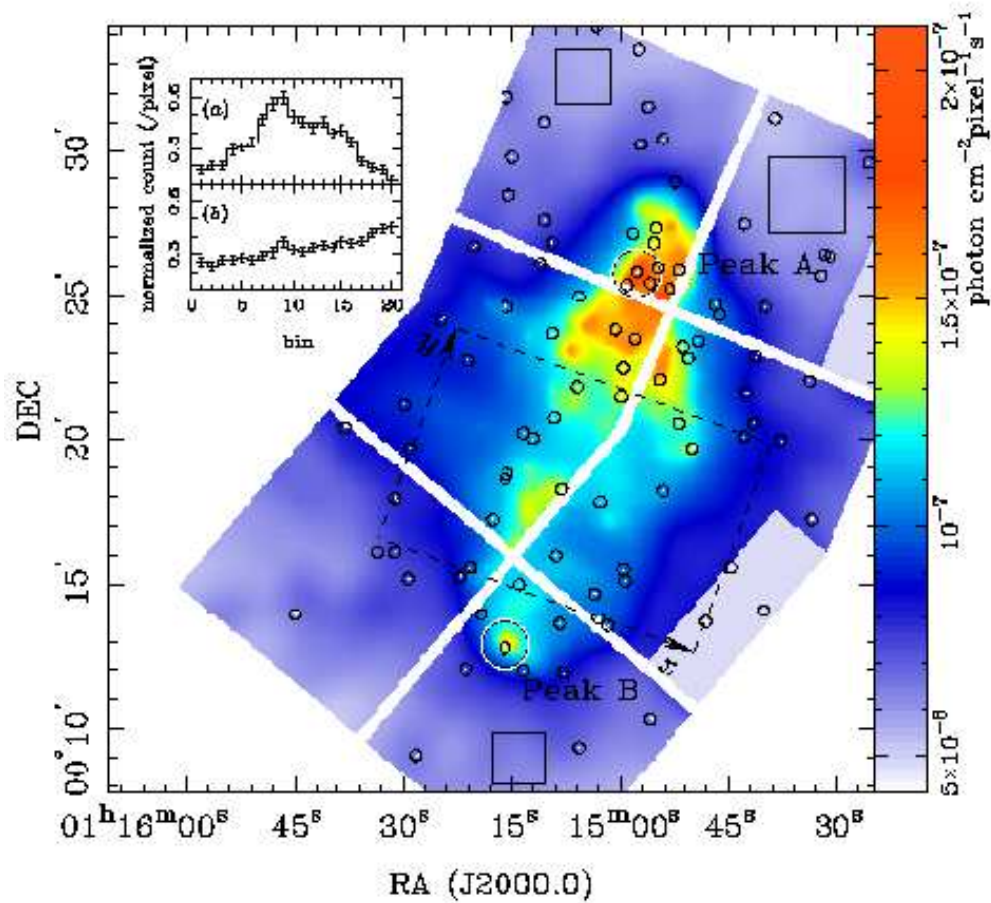}
\caption{
The exposure-map corrected and
adaptively smoothed ACIS-I image of A168 in the 0.3-10 keV band 
(CCD gaps have been excluded). 
The three regions used for estimating 
the mean value and variation of background are shown in boxes. 
The dashed rectangle is used to check the significance of the filament
(see Sect.~\ref{secIMG} for details). The counts along the 
$x$ and $y$ axes of the rectangle are shown in the top left inset,
($a$) and ($b$), respectively. 
The two circles indicate the
apertures used for measuring the flux for the two peaks.
The small circles (with points at the center) indicate removed point sources.
[See the electronic edition of the Journal for a color version of
this figure.]
\label{figIMGxray}}
\end{figure}

\newpage

\begin{figure}
\epsscale{0.8}
\plotone{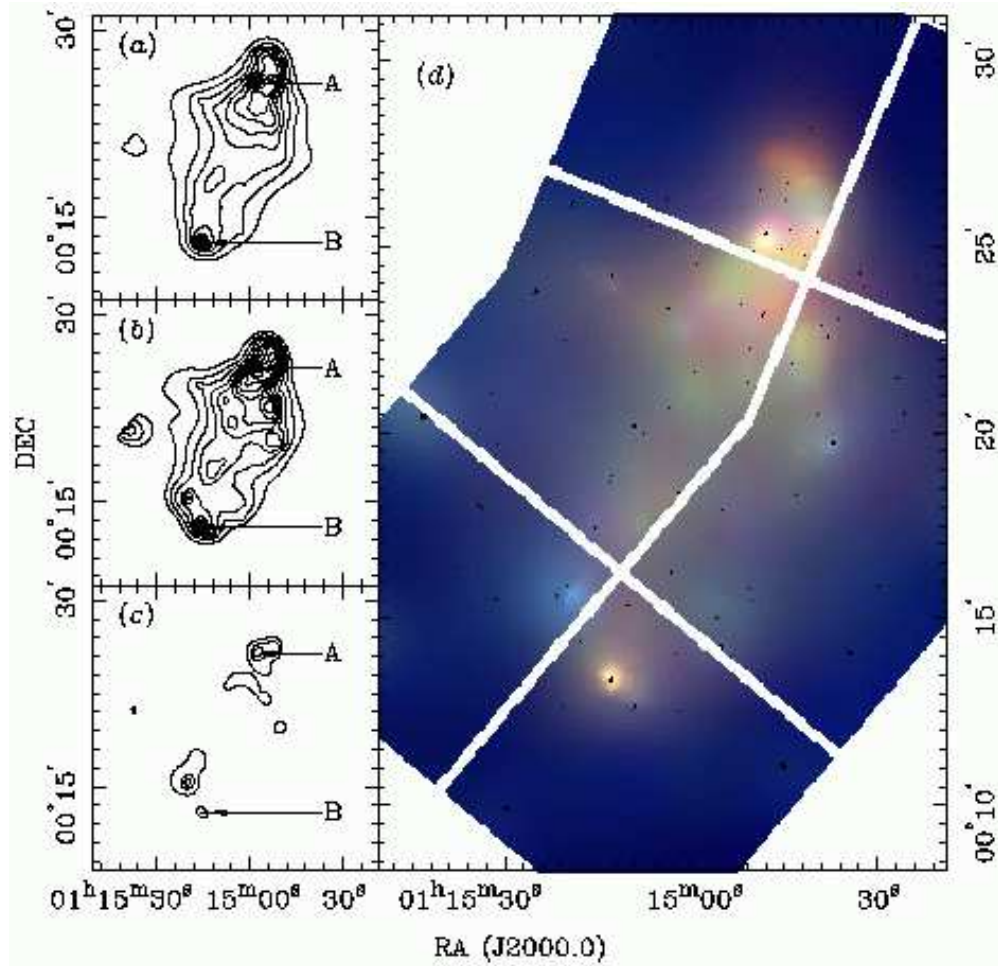}
\caption{
Panel $a$, $b$, $c$ show the contours of soft (0.3-1.5~keV),
medium (1.5-2.5~keV) and hard bands (2.5-10~keV), respectively.
The signal-to-noise ratio for contours in both $a$ and $b$ are 
greater than $5\,\sigma$. 
Contours for panel $c$ vary from $3\,\sigma$ to $8\,\sigma$. 
The two X-ray peaks are labeled as A and B. 
Panel $d$ shows a three-color image created by
combining (linearly) the three band images. 
Redder regions represent the cooler regions; 
bluer regions indicate the hotter regions.
Black points are removed point sources 
(see Figure~\ref{figIMGxray} for clearer view of the point sources).
[See the electronic edition of the Journal for a color version of
this figure.]
\label{figIMGcolor}}
\end{figure}

\newpage

\begin{figure}
\epsscale{0.8}
\plotone{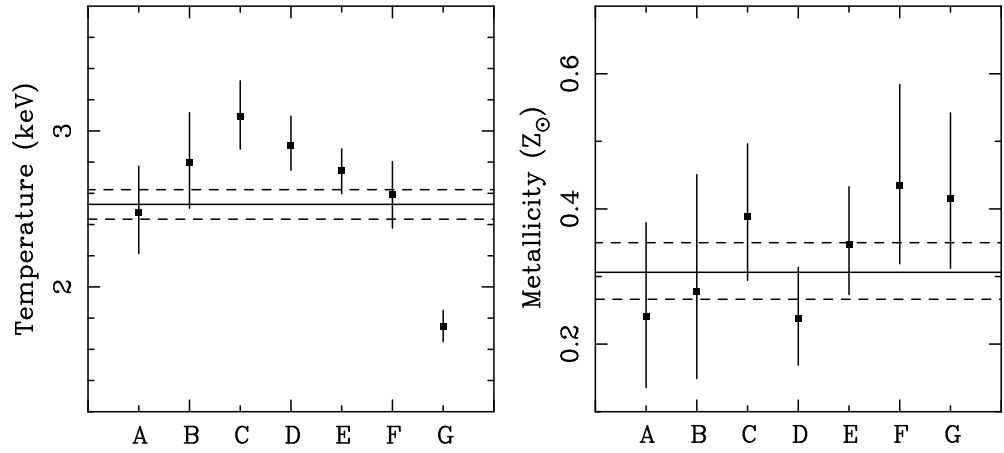}
\caption{
Temperature and metallicity for region A to G defined 
in Figure~\ref{figIMGreg}. 
In each panels, the solid and dashed lines indicate 
the mean value and the 90\% error bars, respectively.
\label{figTA7}}
\end{figure}

\newpage

\begin{figure}
\epsscale{0.8}
\plotone{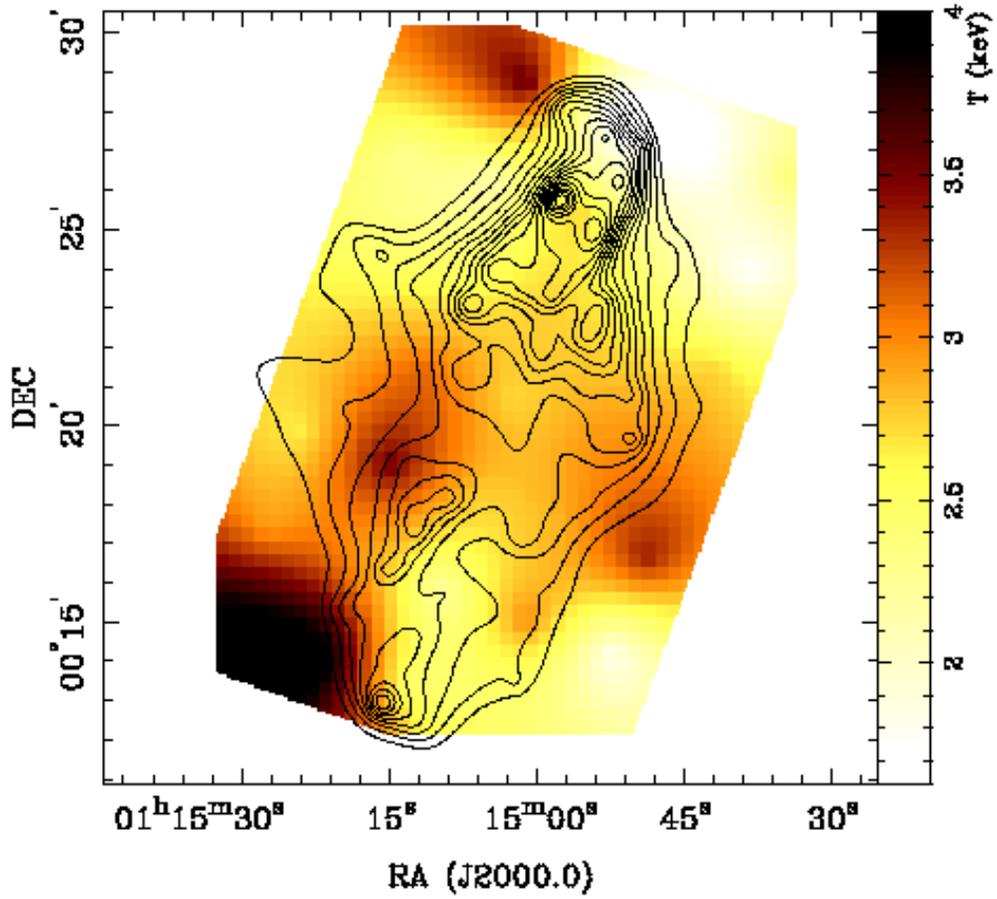}
\caption{Contours ($>3\,\sigma$) of the adaptively smoothed image are
overlayed on the temperature map.
[See the electronic edition of the Journal for a color version of
this figure.]
\label{figTmap}}
\end{figure}

\newpage

\begin{figure}
\epsscale{0.8}
\plotone{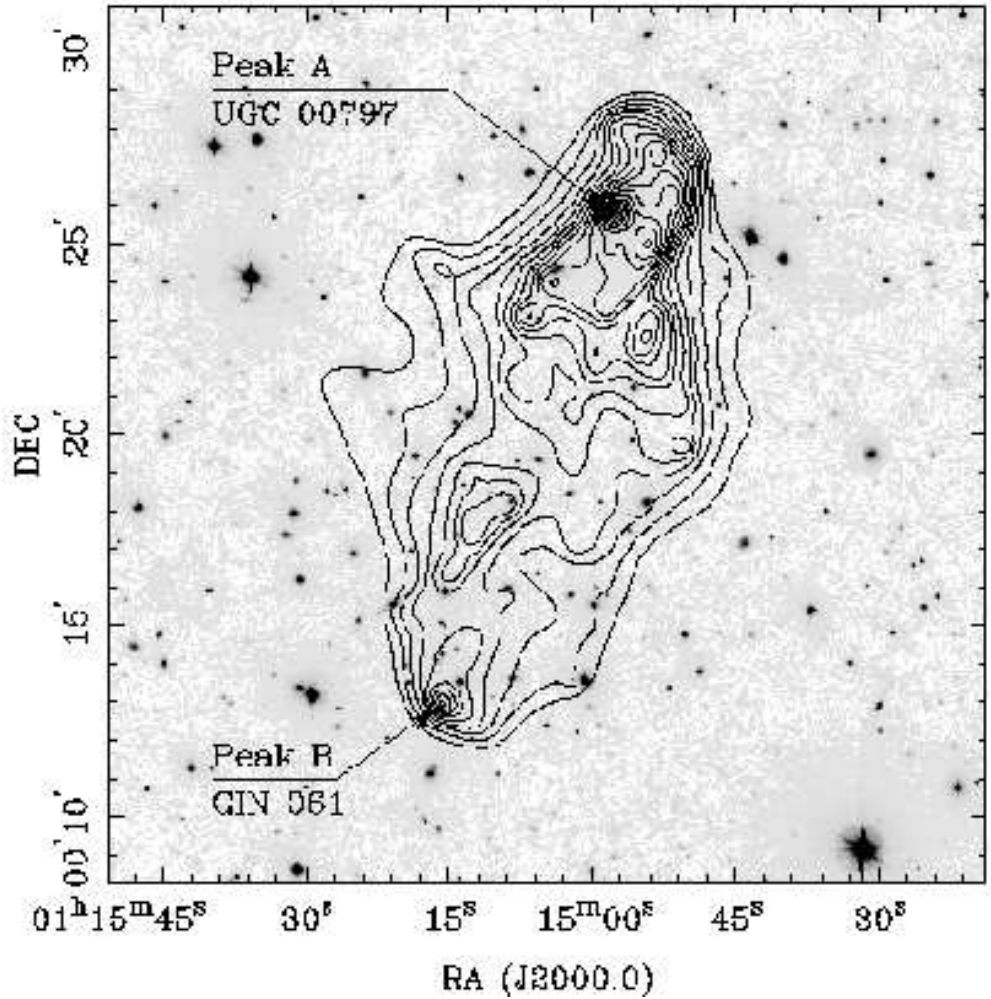}
\caption{
The X-ray contours ($>3\,\sigma$) 
is superposed on the optical   
BATC $i$-band image [BATC is a 15-color photometry program,
see \eg  \citet{1996AJ....112..628F} and \citet{2003A&A...397..361Z}];
the values in the Chandra CCD gaps have been obtained by interpolation. 
The two X-ray peaks are labeled as A and B. 
Peak A is associated with the cD galaxy UGC~797.
Peak B is very close to the second brightest member GIN~061, 
an elliptical galaxy. An elongated filament along the two peaks 
is clearly seen in the contours.
\label{figoptxray}}
\end{figure}

\newpage

\begin{figure}
\epsscale{0.8}
\plotone{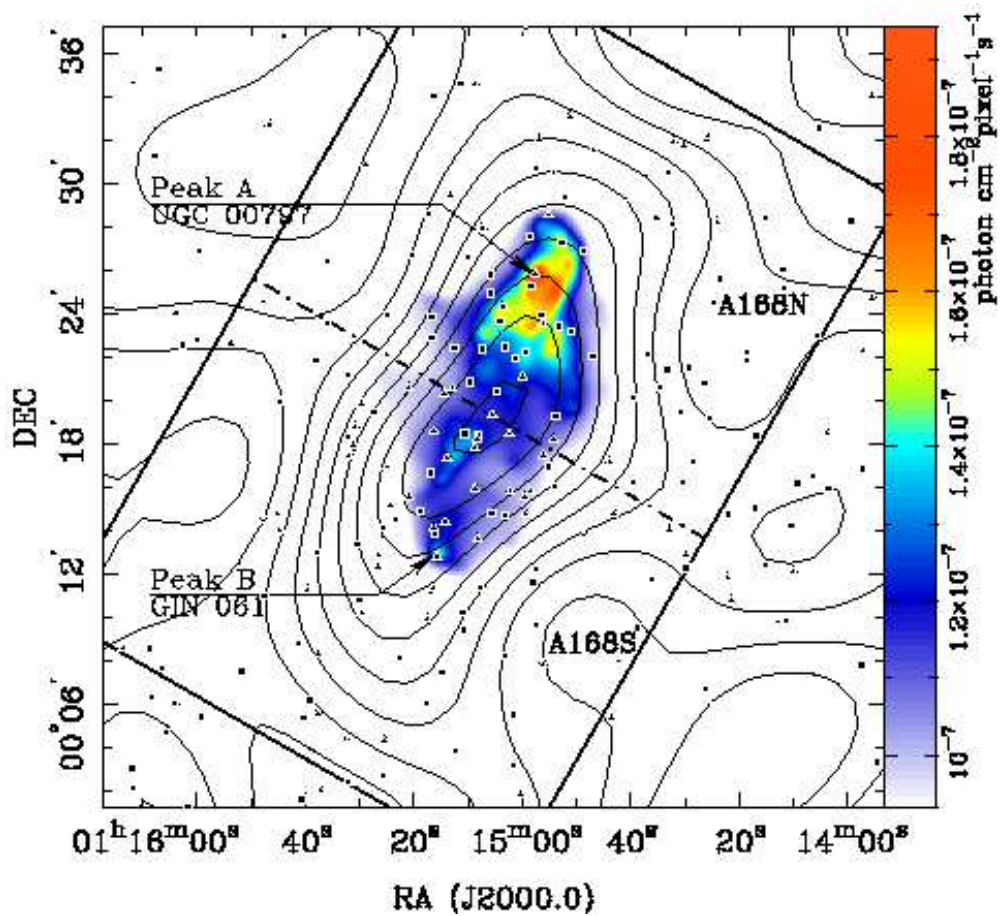}
\caption{
The {\it Chandra} X-ray image is overlayed on the contours
of member galaxies distribution (cf.~Paper\,I). The box (thick-line)
separated by a dotted-dashed line indicates the two subclusters
(A168N and A168S), as suggested in Paper\,I.
The positions of member galaxy are also shown in
the figure: filled-triangles are for the spectroscopically-selected members 
while the filled-squares indicate the members with photometric
redshifts (cf.~Paper\,I). 
[See the electronic edition of the Journal for a color version of
this figure.]
\label{figPaperI}}
\end{figure}


\begin{thebibliography}{}
\bibitem[Albert, White, \& Morgan(1977)]{1977ApJ...211..309A} Albert, C.~E., White, R.~A., \& Morgan, W.~W.\ 1977, \apj, 211, 309 
\bibitem[Bautz \& Morgan(1970)]{1970ApJ...162L.149B} Bautz, L.~P.~\& Morgan, W.~W.\ 1970, \apjl, 162, L149 
\bibitem[Beers, Geller, \& Huchra(1982)]{1982ApJ...257...23B} Beers, T.~C., Geller, M.~J., \& Huchra, J.~P.\ 1982, \apj, 257, 23 
\bibitem[Beers \& Geller(1983)]{1983ApJ...274..491B} Beers, T.~C.~\& Geller, M.~J.\ 1983, \apj, 274, 491 
\bibitem[Bekki(1999)]{1999ApJ...510L..15B} Bekki, K.\ 1999, \apjl, 510, L15 
\bibitem[Condon et al.(1998)]{1998AJ....115.1693C} Condon, J.~J., Cotton, W.~D., Greisen, E.~W., Yin, Q.~F., Perley, R.~A., Taylor, G.~B., \& Broderick, J.~J.\ 1998, \aj, 115, 1693 
\bibitem[David et al.(1993)]{1993ApJ...412..479D} David, L.~P., Slyz, A., Jones, C., Forman, W., Vrtilek, S.~D., \& Arnaud, K.~A.\ 1993, \apj, 412, 479
\bibitem[Dickey \& Lockman(1990)]{1990ARA&A..28..215D} Dickey, J.~M.~\& Lockman, F.~J.\ 1990, \araa, 28, 215
\bibitem[Dressler \& Gunn(1988)]{1988IAUS..130..311D} Dressler, A.~\& Gunn, J.~E.\ 1988, IAU Symp.~130: Large Scale Structures of the Universe, 130, 311 
\bibitem[Fabian et al.(2000)]{2000MNRAS.318L..65F} Fabian, A.~C.~et al.\ 2000, \mnras, 318, L65
\bibitem[Fabian et al.(2003)]{2003MNRAS.344L..43F} Fabian, A.~C., Sanders, J.~S., Allen, S.~W., Crawford, C.~S., Iwasawa, K., Johnstone, R.~M., Schmidt, R.~W., \& Taylor, G.~B.\ 2003, \mnras, 344, L43
\bibitem[Fan et al.(1996)]{1996AJ....112..628F} Fan, X., et al.\ 1996, \aj, 112, 628
\bibitem[Fujita et al.(2002)]{2002ApJ...575..764F} Fujita, Y., Sarazin, C.~L., Kempner, J.~C., Rudnick, L., Slee, O.~B., Roy, A.~L., Andernach, H., \& Ehle, M.\ 2002, \apj, 575, 764 
\bibitem[Gabici \& Blasi(2003)]{2003ApJ...583..695G} Gabici, S.~\& Blasi, P.\ 2003, \apj, 583, 695
\bibitem[Jones et al.(1979)]{1979ApJ...234L..21J} Jones, C., Mandel, E., Schwarz, J., Forman, W., Murray, S.~S., \& Harnden, F.~R.\ 1979, \apjl, 234, L21 
\bibitem[Kaastra \& Mewe(1993)]{1993A&AS...97..443K} Kaastra, J.~S.~\& Mewe, R.\ 1993, \aaps, 97, 443
\bibitem[Liedahl, Osterheld, \& Goldstein(1995)]{1995ApJ...438L.115L} Liedahl, D.~A., Osterheld, A.~L., \& Goldstein, W.~H.\ 1995, \apjl, 438, L115
\bibitem[Markevitch, Vikhlinin, \& Mazzotta(2001)]{2001ApJ...562L.153M}Markevitch, M., Vikhlinin, A., \& Mazzotta, P.\ 2001, \apjl, 562, L153 
\bibitem[Markevitch \& Vikhlinin(2001)]{2001ApJ...563...95M} Markevitch, M.~\& Vikhlinin, A.\ 2001, \apj, 563, 95
\bibitem[Markevitch et al.(2002)]{2002ApJ...567L..27M} Markevitch, M., Gonzalez, A.~H., David, L., Vikhlinin, A., Murray, S., Forman, W., Jones, C., \& Tucker, W.\ 2002, \apjl, 567, L27
\bibitem[Markevitch et al.(2003)]{2003ApJ...583...70M} Markevitch, M., et al.\ 2003, \apj, 583, 70 
\bibitem[Marvel, Shukla, \& Rhee(1999)]{1999ApJS..120..147M} Marvel, K.~B., Shukla, H., \& Rhee, G.\ 1999, \apjs, 120, 147 
\bibitem[Mihos(2004)]{2004cgpc.symp..278M} Mihos, J.~C.\ 2004,
in Carnegie Observatories Astrophysics Series, Vol. 3: Clusters of
Galaxies: Probes of Cosmological Structure and Galaxy Evolution,
ed. J. S. Mulchaey, A. Dressler, \& A. Oemler (Cambridge: Cambridge
Univ. Press), p.~278 (astro-ph/0305512)
\bibitem[Morgan \& Lesh(1965)]{1965ApJ...142.1364M} Morgan, W.~W.~\& Lesh, J.~R.\ 1965, \apj, 142, 1364 
\bibitem[Morgan, Kayser, \& White(1975)]{1975ApJ...199..545M} Morgan, W.~W., Kayser, S., \& White, R.~A.\ 1975, \apj, 199, 545 
\bibitem[Oegerle \& Hill(2001)]{2001AJ....122.2858O} Oegerle, W.~R.~\& Hill, J.~M.\ 2001, \aj, 122, 2858 
\bibitem[Oilver \& Webster(1990)]{1990IJGICS...4..313O} Oliver, M. A.,
\& Webster, R. 1990, INT. J. Geographical Information Systems, 4, 313
\bibitem[Ricker(1998)]{1998ApJ...496..670R} Ricker, P.~M.\ 1998, \apj, 496, 670
\bibitem[Ricker \& Sarazin(2001)]{2001ApJ...561..621R} Ricker, P.~M.~\& Sarazin, C.~L.\ 2001, \apj, 561, 621
\bibitem[Roettiger, Loken, \& Burns(1997)]{1997ApJS..109..307R} Roettiger, K., Loken, C., \& Burns, J.~O.\ 1997, \apjs, 109, 307     
\bibitem[Roettiger, Stone, \& Burns(1999)]{1999ApJ...518..594R} Roettiger, K., Stone, J.~M., \& Burns, J.~O.\ 1999, \apj, 518, 594                         
\bibitem[Sun et al.(2002)]{2002ApJ...565..867S} Sun, M., Murray, S.~S., Markevitch, M., \& Vikhlinin, A.\ 2002, \apj, 565, 867
\bibitem[Sun \& Murray(2002)]{2002ApJ...576..708S} Sun, M.~\& Murray, S.~S.\ 2002, \apj, 576, 708 
\bibitem[Tomita et al.(1996)]{1996AJ....111...42T} Tomita, A., Nakamura, F.~E., Takata, T., Nakanishi, K., Takeuchi, T., Ohta, K., \& Yamada, T.\ 1996, \aj, 111, 42
\bibitem[Ulmer, Wirth, \& Kowalski(1992)]{1992ApJ...397..430U} Ulmer,M.~P., Wirth, G.~D., \& Kowalski, M.~P.\ 1992, \apj, 397, 430
\bibitem[Vikhlinin, Markevitch, \& Murray(2001)]{2001ApJ...551..160V} Vikhlinin, A., Markevitch, M., \& Murray, S.~S.\ 2001, \apj, 551, 160
\bibitem[White, Jones, \& Forman(1997)]{1997MNRAS.292..419W} White, D.~A., Jones, C., \& Forman, W.\ 1997, \mnras, 292, 419
\bibitem[Yang et al.(2004)]{2004ApJ...600..141Y} Yang, Y., Zhou, X., Yuan, Q., Jiang, Z., Ma, J., Wu, H., \& Chen, J.\ 2004, \apj, 600, 141
\bibitem[Zhou et al.(2003)]{2003A&A...397..361Z} Zhou, X., et al.\ 2003, \aap, 397, 361
\end{thebibliography}
\end{document}